\documentclass[aps,showpacs,preprintnumbers]{revtex4}

\usepackage{amsfonts}

\usepackage{graphicx}% Include figure files
\usepackage{dcolumn}% Align table columns on decimal point
\usepackage{bm}% bold math
\usepackage{epsfig}
\newcommand{\vsig}{\mbox{\boldmath$\sigma$\unboldmath}}

\begin{document}

\title{Strong decays of newly observed $D_{sJ}$ states in a constituent quark model with effective Lagrangians}
\author{
Xian-Hui Zhong$^{1,3}$\footnote {E-mail: zhongxh@ihep.ac.cn} and
Qiang Zhao$^{2,3}$\footnote {E-mail: zhaoq@ihep.ac.cn}}

\affiliation{1)  Department of Physics, Hunan Normal University, and
Key Laboratory of Low-Dimensional Quantum Structures $\&$ Quantum
Control of Ministry of Education, Changsha 410081, P.R. China }

\affiliation{2) Institute of High Energy Physics,
       Chinese Academy of Sciences, Beijing 100049, P.R. China
}

\affiliation{3) Theoretical Physics Center for Science Facilities,
CAS, Beijing 100049, P.R. China}

%\date{\today}

\begin{abstract}
The strong decay properties of the newly observed states
$D_{sJ}(3040)$, $D_{sJ}(2860)$ and $D_{sJ}(2710)$ are studied in a
constituent quark model with quark-meson effective Lagrangians. We
find that the $D_{sJ}(3040)$ could be identified as the low mass
physical state $|2{P_1}\rangle_L$ ($J^P=1^+$) from the
$D_{s}(2^1P_1)$-$D_{s}(2^3P_1)$ mixing. The $D_{sJ}(2710)$ is likely
to be the low-mass mixed state $|(SD)\rangle_L$ via the
$1^3D_1$-$2^3S_1$ mixing. In our model, the $D_{sJ}(2860)$ cannot be
assigned to any single state with a narrow width and compatible
partial widths to $DK$ and $D^*K$. Thus, we investigate a two-state
scenario as proposed in the literature. In our model, one resonance
is likely to be the $1^3D_3$ ($J^P=3^-$), which mainly decays into
$DK$. The other resonance seems to be the $|1{D_2}'\rangle_H$, i.e.
the high-mass state in the $1^1D_2$-$1^3D_2$ mixing with $J^P=2^-$,
of which the $D^*K$ channel is its key decay mode. We also discuss
implications arising from these assignments and give predictions for
their partner states such as $|(SD)'\rangle_H$, $|2{P'_1}\rangle_H$,
$2^3P_0$ and $2^3P_2$, which could be helpful for the search for
these new states in future experiment.

\end{abstract}


\maketitle

\section{Introduction}

Experimental progress on the study of $D$ and $D_s$ states in the
past few years provides a great opportunity for theory development.
Recently a new broad resonance $D_{sJ}(3040)$ with a mass of
$(3044\pm 8_{\mathrm{stat}}(^{+30}_{-5})_\mathrm{syst})$ MeV and a
width of $\Gamma=(239\pm
35_{\mathrm{stat}}(^{+46}_{-42})_\mathrm{syst})$ MeV is reported in
the $D^*K$ channel \cite{Aubert:2009}. Apart from the $D_{sJ}(3040)$
another two states $D_{sJ}(2710)$ and $D_{sJ}(2860)$, which were
observed by BABAR and Belle two years
ago~\cite{Aubert:2006mh,jb:2007aa}, are also examined. Their
branching ratio fractions between $D^*K$ and $DK$ are measured
\cite{Aubert:2009},
\begin{eqnarray}
\frac{D_{sJ}(2710)^+\rightarrow D^{*} K}{D_{sJ}(2710)^+\rightarrow
DK}=0.91\pm0.13_{\mathrm{stat}}\pm 0.12_{\mathrm{syst}},\\
\frac{D_{sJ}(2860)^+\rightarrow D^{*} K}{D_{sJ}(2860)^+\rightarrow
DK}=1.10\pm0.15_{\mathrm{stat}}\pm 0.19_{\mathrm{syst}}.
\end{eqnarray}
These new observations stimulate great interest in the understanding
of their nature and strong coupling properties in theory. Different
theoretical approaches for the study of the strong coupling
properties of heavy-light mesons can be found in the literature,
such as the heavy quark effective field theory approach
(HQEFT)~\cite{Nowak:1992um,Deandrea:1999pa,Deandrea:1998uz,Dai:1997df,Eichten:1993ub,
Dai:1998ve,Hiorth:2002pp,Fajfer:2006hi,Fajfer:2008zz,
 Kamenik:2008zza,Bardeen:1993ae,Bardeen:2003kt,Colangelo:2004vu,Colangelo:2007ds,Fazio}, QCD
sum rules \cite{Colangelo:1995ph,Dai:1997df,Wang:2008tm}, $^3P_0$
model
\cite{Close:2006gr,Close:2005se,Zhang:2006yj,Wei:2006wa,Liuxiang},
and chiral quark model \cite{Pierro}.

In this work, we present an analysis of these $D_s$ states in a
constituent quark model with effective Lagrangians for the
quark-meson couplings, and try to clarify the following issues: (i)
To gain information about the structure of the newly observed state
$D_{sJ}(3040)$ according to its strong decay properties. (ii) With
the new data for the $D_{sJ}(2710)$ and $D_{sJ}(2860)$, we reanalyze
their strong decays and examine their structures again. The quantum
numbers of these two states remain controversial. The $D_{sJ}(2710)$
is identified as a state of $J^P=1^-$ in $B$ decays
\cite{jb:2007aa}, while it is explained by various models as the
$2^3S_1$, $1^3D_1$, the admixtures of $2^3S_1$-$1^3D_1$, molecular
structure, or tetraquark state
\cite{Zhang:2006yj,Wei:2006wa,Close:2006gr,Colangelo:2007ds,Wang:2007nfa}.
There are also a lot of solutions proposed for the $D_{sJ}(2860)$.
The assignments of $1^3D_3$ or $2^3P_0$ have been discussed in Refs.
\cite{Zhang:2006yj,Close:2006gr,vanBeveren:2006st,Wei:2006wa,Colangelo:2006rq,Koponen:2007nr,Zhong:2008kd}.
A recent comment by Ref. \cite{eef:2009} suggests a two-state
structure for the $D_{sJ}(2860)$ in order to understand the
controversial aspects arising from its strong decays. (iii) The
quark-model assignment of these states will result in implications
of their partner multiplets. We discuss some of those relevant
states, for which the experimental observations would be able to
clarify some of those theoretical and experimental issues.

By treating the light mesons (pseudoscalar and vector mesons) as
effective fields, we introduce constituent-quark-meson couplings to
describe the charmed meson strong decays into a charmed meson plus a
light pseudoscalar or vector meson in the final state. The
quark-pseudoscalar-meson coupling is given by the chiral quark model
at the leading order as proposed by Manohar and
Georgi~\cite{Manohar:1983md}. Its application to pseudoscalar meson
photoproduction in the quark model turns out to be promising and
many low-energy phenomena can be highlighted in such a
framework~\cite{Li:1994cy,Li:1997gda,qk3,Zhong:2007fx,Zhong:2009}.
In particular, the axial current conservation allows one to extract
the axial coupling in terms of the meson decay constant and a form
factor arising from the microscopic quark model
wavefunctions~\cite{Riska:2000gd}. With an effective
quark-vector-meson coupling, one can also extract the vector
couplings in a similar
way~\cite{zhao:1998fn,zhao-k1,zhao-kstar,Riska:2000gd}.

A natural extension of this picture is to apply this effective
Lagrangian approach to heavy-light meson strong decays involving
light pseudoscalar or vector mesons, which would be a good place to
examine the validity of the light axial and vector fields in such a
transition. On the one hand, the quark-meson coupling is the same as
that defined in meson photoproduction which is proportional to the
meson decay constant. On the other hand, the heavy-light meson in
the initial and final state would provide information about the
coupling form factor and can be calculated in the quark model
framework. Thus, one can study the heavy-light meson strong decays
by combining dynamical information from meson photoproduction off
nucleons.

The paper is organized as follows. In Sec.\ \ref{suc}, a brief
review of the quark-meson effective Lagrangian approach is given.
The numerical results are presented and discussed in Sec.\
\ref{cpb}. Finally, a summary is given in Sec.\ \ref{suma}.

\section{framework}\label{suc}

\begin{center}
\begin{figure}[ht]
\centering \epsfxsize=8 cm \epsfbox{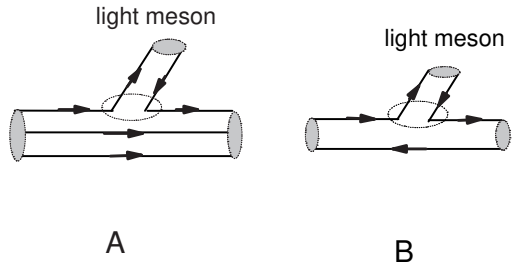} \caption{ Diagrams
(A) and (B) stand for the quark-meson couplings in meson-baryon
interactions and light-meson production in heavy-light meson strong
decays, respectively.}\label{fig-q}
\end{figure}
\end{center}

In Fig.~\ref{fig-q}, we illustrate the similarity of the quark-meson
coupling in meson-baryon and light-meson production in heavy-light
meson strong decays. It should be pointed out that since the flavor
symmetries beyond the SU(3) are badly broken, the contributions from
transitions of treating the final-state heavy-light meson as an
effective field are strongly suppressed. We thus can neglect those
contributions safely in our approach. An early study of the charmed
meson strong decays can be found in Ref.~\cite{Zhong:2008kd}.

In the chiral quark model~\cite{Manohar:1983md}, the low energy
quark-pseudoscalar-meson interactions in the SU(3) flavor basis are
described by the effective
Lagrangian~\cite{Li:1997gda,qk3,qk4,Zhong:2007fx,Zhong:2009}
\begin{equation}\label{coup}
{\cal L}_{Pqq}=\sum_j
\frac{1}{f_m}\bar{\psi}_j\gamma^{j}_{\mu}\gamma^{j}_{5}\psi_j\partial^{\mu}\phi_m.
\end{equation}
where $\psi_j$ represents the $j$-th quark field in the hadron, and
$\phi_m$ is the pseudoscalar meson field.

The effective Lagrangian for quark-vector-meson interactions in the
SU(3) flavor basis is~\cite{zhao:1998fn,zhao-k1,zhao-kstar}
\begin{equation}
{\cal
L}_{Vqq}=\sum_j\bar{\psi}_j(a\gamma^{j}_{\mu}+\frac{ib}{2m_j}\sigma_{\mu\nu}q^\nu)V^\mu\psi_j
\ ,
\end{equation}
where $V^\mu$ represents the vector meson field with four-vector
moment $q$. Parameters $a$ and $b$ denote the vector and tensor
coupling strength, respectively.

As follows, we provide the quark-pseudoscalar and quark-vector-meson
coupling operators in a non-relativistic
form~\cite{Li:1997gda,qk3,qk4,Zhong:2007fx,Zhong:2009,zhao:1998fn,zhao-k1,zhao-kstar}.
Considering light meson emission in a heavy-light meson strong
decays, the effective quark-pseudoscalar-meson coupling operator in
the center-of-mass (c.m.) system of the initial meson is
\begin{eqnarray}\label{ccpk}
H_{m}=\sum_j\left[-\left(1+\frac{\omega_m}{E_f+M_f}\right)\vsig_j
\cdot \textbf{q} +\frac{\omega_m}{2\mu_q}\vsig_j\cdot
\textbf{p}_j\right]I_j \varphi_m \ .
\end{eqnarray}

In a case that a light vector meson is emitted, the transition
operators for producing a transversely or longitudinally polarized
vector meson are as follows
\begin{eqnarray}
H_m^T=\sum_j \left\{i\frac{b'}{2m_q}\vsig_j\cdot
(\mathbf{q}\times\mathbf{\epsilon})+\frac{a}{2\mu_q}\mathbf{p}_j\cdot
\mathbf{\epsilon}\right\}I_j\varphi_m,
\end{eqnarray}
and
\begin{eqnarray}
H_m^L=\sum_j \frac{a M_v}{|\mathbf{q}|}I_j\varphi_m \ .
\end{eqnarray}
In the above three equations,  $\textbf{q}$ and $\omega_m$ are the
three-vector momentum and energy of the final-state light meson,
respectively. $\textbf{p}_j$ is the internal momentum operator of
the $j$-th quark in the heavy-light meson rest frame. $\vsig_j$ is
the spin operator on the $j$-th quark of the heavy-light system and
$\mu_q$ is a reduced mass given by $1/\mu_q=1/m_j+1/m'_j$ with $m_j$
and $m'_j$ for the masses of the $j$-th quark in the initial and
final mesons, respectively. Here, the $j$-th quark is referred to
the active quark involved at the quark-meson coupling vertex. $M_v$
is the mass of the emitted vector meson. The plane wave part of the
emitted light meson is $\varphi_m=e^{-i\textbf{q}\cdot
\textbf{r}_j}$, and $I_j$ is the flavor operator defined for the
transitions in the SU(3) flavor space
\cite{Zhong:2007gp,Li:1997gda,qk3,qk4,Zhong:2009,Zhong:2007fx,Zhong:2008kd,zhao:1998fn}.
Parameters $a$ and $b$ are the vector and tensor coupling strengths
of the quark-vector-meson couplings, respectively. Studies of vector
meson photoproduction
\cite{zhao-k1,zhao-kstar,JLab-Kstar,elsa-kstar} suggest that
$a=g_{\omega qq}=g_{\rho qq}\simeq -3$ and $b'\equiv b-a \simeq 5$.
Because of vector current conservation, one has $a=g_{\rho
NN}=g_{\omega NN}/3$ \cite{Riska:2000gd,zhao:1998fn,zhao-kstar}.

The heavy-light meson wavefunctions have been given in Ref.
\cite{Zhong:2008kd}, and some of the decay amplitudes have also been
deduced there. In the charmed meson decays, the SU(4) flavor
symmetry is broken. Thus, the charm quark is treated as a spectator
and the transition amplitude is proportional to the final-state
light meson decay constant associated with a form factor arising
from the convolution of the initial and final-state charmed meson
wavefunctions.

In the calculation, the standard quark model parameters are adopted.
Namely, we set $m_u=m_d=330$ MeV, $m_s=450$ MeV, and $m_c=1700$ MeV
for the constituent quark masses. The harmonic oscillator parameter
$\beta$ is usually adopted in the range of (0.4--0.5) GeV, in this
work we take it as $\beta=0.45$ GeV. The decay constants for $K$ and
$\eta$ mesons are $f_K=f_{\eta}=160$ MeV. As shown in Refs.
\cite{Zhong:2007gp,Zhong:2008kd}, the flavor symmetry breaking will
lead to corrections to the quark-pseudoscalar-meson coupling vertex,
for which an additional global parameter $\delta$ is introduced.
Here, we fix its value the same as that in
Refs.~\cite{Zhong:2008kd,Zhong:2007gp}, i.e. $\delta=0.557$. For the
quark-vector-meson coupling strength which still suffers relatively
large uncertainties, we adopt the values extracted from vector meson
photoproduction as mentioned earlier, i.e. $a\simeq -3$ and
$b'\simeq 5$. The masses of the mesons used in the calculations are
adopted from the PDG~\cite{PDG}.

Justification of the non-relativistic formulation is not obvious for
the light quark sector in the heavy-light meson transitions. This is
similar to the case of a non-relativistic quark model for baryons,
where the results would rely on the experimental data to tell how
far they deviate from reality. Treating the light meson as a chiral
field somehow assumes that the light meson is produced at short
distance, and the spectators (i.e. the two spectator quarks inside a
baryon or the heavy quark in the heavy-light meson transitions) do
not respond to the internal structure of the light meson. Instead,
the propagation of the light quark pair would feel the hadronic
environment from the convolution of initial and final-state
heavy-light mesons. Such an implicated assumption means that only
the processes with relatively small momentum transfers between the
light quarks inside the light meson would dominantly contribute to
the transition matrix element. This empirically supports the
validity of the non-relativistic formulation as a leading order
approximation.

\section{results and discussions}\label{cpb}

\subsection{$D_{sJ}(3040)$}

The $D_{sJ}(3040)$ is observed in the $D^*K$ mode, while there is no
sign of $D_{sJ}(3040)\rightarrow DK$ in experiment
\cite{Aubert:2009}. This allows its quantum number to be $J^P=0^-$,
$1^+$, $2^-$ etc. The $J^P=0^-$ state $2^1S_0$ seems not a good
candidate since its predicted mass, $\sim 2.7$ GeV
\cite{Pierro,Godfrey:1985xj,Close:2006gr,Ebert:2009ua}, is much less
than $3.04$ GeV. The  predicted masses of $J^P=1^+$ and $2^-$ are
close to $3.04$ GeV. We hence discuss these two possibilities for
the $D_{sJ}(3040)$ in this work.

First, we considered it as the $J^P=1^+$ states $2^1P_1$ and
$2^3P_1$. These two states also can decay into $D^*K$, $DK^*$,
$D^*_s\eta$, $D_s\phi$, $D_0(2400)K$, $D_1(2430)K$, $D_1(2420)K$,
$D_2(2460)K$, $D_s(2317)\eta$, $D_s(2460)\eta$. With a mass of 3.04
GeV, we calculate their decay widths, which are listed in Tab.
\ref{wf2}. From the table, it is found that the decay width of
$2^1P_1$ and $2^3P_1$ are $\sim 115$ MeV and $\sim93$ MeV,
respectively, which are too small to compare with the data, although
the decay mode, dominated by the $D^*K$, is consistent with the
observation~\cite{Aubert:2009}. Thus, the $D_{sJ}(3040)$ may not be
considered as pure $2^1P_1$ or $2^3P_1$ state.

\begin{center}
\begin{figure}[ht]
\centering \epsfxsize=10 cm \epsfbox{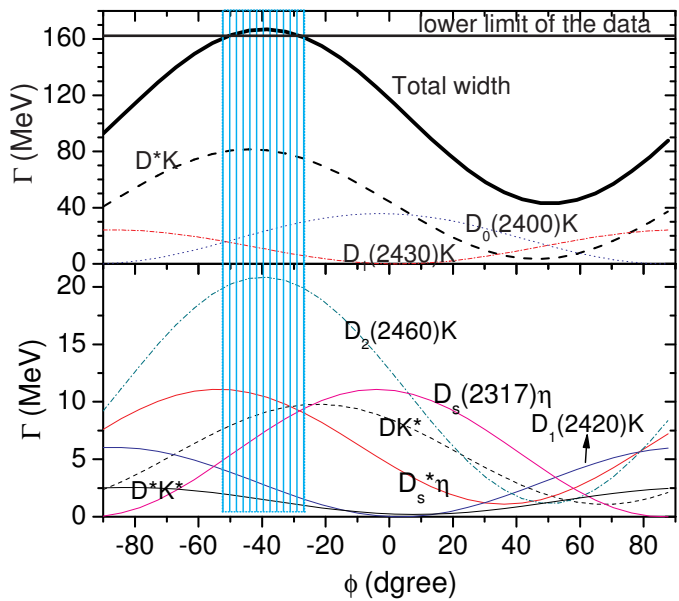} \caption{ (Color
online) The partial decay widths and total width of $|2P_1\rangle_L$
with a mass of 3040 MeV as functions of mixing angle $\phi$. The
data are from BABAR \cite{Aubert:2009}.}\label{fig-1}
\end{figure}
\end{center}

Since the heavy-light mesons are not charge conjugation eigenstates,
state mixing between spin $\mathbf{S}=0$ and $\mathbf{S}=1$ states
with the same $J^P$ can occur via  the spin-orbit interactions
\cite{Godfrey:1986wj,Close:2005se,Swanson}. The physical states with
$J^P=1^-$ can then be described as
\begin{eqnarray}
|2P_1\rangle_L=+\cos (\phi) |2^1P_1\rangle+\sin(\phi)|2^3P_1\rangle, \\
|2{P_1}'\rangle_H=-\sin (\phi)
|2^1P_1\rangle+\cos(\phi)|2^3P_1\rangle \ ,
\end{eqnarray}
where the subscripts $L$ and $H$ stand for the low mass and high
mass of the physical states after the mixing.

Usually, the low mass state has a broad width while the high mass
state has a narrow width. We set the mass of $|2P_1\rangle_L$ with
$3.04$ GeV, and plot its decay width as a function of the mixing
angle $\phi$, which is shown in Fig.~\ref{fig-1}. It shows that when
the mixing angle is in the range $\phi\simeq-(40\pm 12)^\circ$, the
total decay width, $\Gamma=(162\sim 170)$ MeV, is in the range of
the experimental data (close to the lower limit of the
data)~\cite{Aubert:2009}. The mixing angle predicted here is
consistent with the result $\phi\simeq -55^\circ$ in the heavy quark
limit~ \cite{Godfrey:1986wj,Close:2005se,Swanson}. The $D^*K$
governs the decays of $|2P_1\rangle_L$, while the $DK$ channel is
forbidden. This is also in agreement with the observations. These
results suggest that the $D_{sJ}(3040)$ favors the $|2P_1\rangle_L$
classification.

Apart from the $D^*K$ mode, the $D_1(2430)K$, $D_2(2460)K$,
$D_0(2400)K$, $DK^*$, and $D_s^*\eta$ are also important in the
decays of $|2P_1\rangle_L$ as shown by Fig.~\ref{fig-1}. In
particular, the partial widths of $D_1(2420)K$, $D_s(2317)\eta$ and
$D^*K^*$ turn out to be sizable. A search for those channels would
be useful for clarifying the property of the $D_{sJ}(3040)$. With
the mixing angle $\phi\simeq -55^\circ$, the relative decay ratios
among those decay channels are $D^*K:D_1(2430)K
:D_2(2460)K:D_0(2400)K:D_1(2420)K:DK^*:D_s(2317)\eta:D_s^*\eta:D^*K^*=78:17:19:13:4:8:4:11:2$.

\begin{center}
\begin{figure}[ht]
\centering \epsfxsize=10 cm \epsfbox{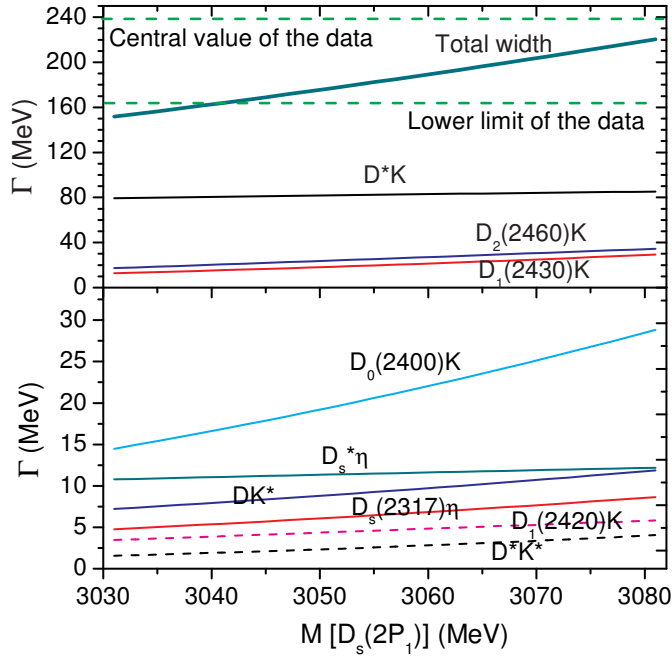} \caption{ (Color
online) The partial decay widths and total width of $|2P_1\rangle_L$
as functions of mass. The data are from BABAR \cite{Aubert:2009}.
}\label{fig-2}
\end{figure}
\end{center}

Since the mass of $D_s(3040)$ still has a large uncertainty, it may
bring uncertainties to the theoretical predictions on the decay
widths. To investigate this effect, we plot the decay widths as a
function of the mass in Fig. \ref{fig-2} with the mixing angle fixed
at $\phi=-50^\circ$. It shows that the mass uncertainty gives rise
to an uncertainty of about $\sim 70$ MeV in the total decay width.
The predicted widths are much closer to the central value of the
data with the increasing mass. The sensitivity of different decay
modes to the mass can also be seen clearly in the plot.

\begin{table}[ht]
\caption{The decay widths (MeV) for the $D_{sJ}(3040)$ as $1^1D_2$,
$1^3D_2$, $2^1P_1$ and $2^3P_1$ candidates.} \label{wf2}
\begin{tabular}{|c|c|c|c|c|c|c |c|c|c|c|c|c|c|c}\hline\hline
& $D^*K$ & $DK^* $ & $D^{*}K^*$ &$D^*_s\eta$& D(2430)K & D(2420)K &
$D_s(2460)\eta$
& $D_s\phi$ & D(2400)K & D(2460)K & $D_s(2317)\eta$ & total   \\
\hline
$1^1D_2$ &197     & 27 &2   & 25 & 3  & 2 & 0.4 &4 & 3 &345 & 4 & 608    \\
\hline
$1^3D_2 $ &256    & 21 & 33   &34  & 1 & 18 &0.01 & 0.05 & 3.4 & 512 &1.6& 879\\
\hline
$2^1P_1 $ &44    & 9 & 0.3   &5.5  & 0.02 & 0.01 &$7.5\times10^{-5}$ & 0.1 & 33 & 12 &11& 115\\
\hline
$2^3P_1 $ &41    & 2 & 2.5   &7.5  & 24 & 7 &0.5 & 0.002 & 0.09 & 9 &0.06& 93\\
\hline
\end{tabular}
\end{table}

Finally, we discuss the possibilities of $D_{sJ}(3040)$ as a
$J^P=2^-$ candidate. There are two states, $1^1D_2$ and $1^3D_2$,
with $J^P=2^-$. If $1^1D_2$ and $1^3D_2$ have a mass of 3.04 GeV,
they can decay into the following channels, $D^*K$, $DK^*$,
$D^*K^*$, $D^*_s\eta$, $D_s\phi$, $D_0(2400)K$, $D_1(2430)K$,
$D_1(2420)K$, $D_2(2460)K$, $D_s(2317)\eta$, and $D_s(2460)\eta$. We
calculate these partial decay widths and list the results in Tab.
\ref{wf2}. It shows that $D^*K$ and $D_2(2460)K$ are the two main
decay channels. The total widths for both $1^1D_2$ and $1^3D_2$ are
very broad, i.e. $\Gamma\sim 608$ MeV and $\sim 879$ MeV,
respectively. They are too large to compare with the data
$\Gamma=(239\pm 35)$ MeV~\cite{Aubert:2009}. Nevertheless, it shows
that the admixtures between $1^1D_2$ and $1^3D_2$ are unable to give
a reasonable explanation of the decay properties of $D_{sJ}(3040)$
as well. Thus, the $D_{sJ}(3040)$ as a $J^P=2^-$ candidate is not
favored.

In brief, the $D_{sJ}(3040)$ seems to favor a $|2P_1\rangle_L$ state
with $J^P=1^+$, which is an admixture of $2^1P_1$ and $2^3P_1$ with
a mixing angle $\phi\simeq-(40\pm 12)^\circ$. Our conclusion is in
agreement with that of a $^3P_0$ model analysis~\cite{Liuxiang}. The
semiclassical flux tube model~\cite{Ailin} and relativistic quark
model~\cite{Ebert:2009ua} mass calculations also support this
picture.

\subsection{$D_{sJ}(2710)$} \label{cy}

The $D_{sJ}(2710)$ was first reported by BABAR~\cite{Aubert:2006mh},
and its quantum number $J^P=1^-$ was determined by
Belle~\cite{jb:2007aa}. Recently, the decay ratios of the
$D_{sJ}(2710)$ have also been reported~\cite{Aubert:2009}, which is
very useful for understanding its nature. According to the
classification of the quark model, only two states $2^3S_1$ and
$1^3D_1$ with the quantum number $J^P=1^-$ are located around the
mass range $(2.7\sim 2.8)$ GeV. This state is studied by various
models, e.g. as a $2^3S_1$ state~\cite{Ebert:2009ua,Fazio}, $1^3D_1$
state \cite{Wei:2006wa}, or admixture of $2^3S_1$-$1^3D_1$
\cite{Close:2006gr}.  It should mention that in our previous work
\cite{Zhong:2008kd} an error occurred in the partial decay amplitude
of $1^3D_1\rightarrow DK$, which led to a rather small width for the
assignment of the admixture of $2^3S_1$-$1^3D_1$. Here we correct
the formulation and reanalyze the mixing scenario for the
$D_{sJ}(2710)$.

\begin{table}[ht]
\caption{The decay widths (MeV) for the $D_{sJ}(2710)$ as $1^3D_1$
and $2^3S_1$ candidates.} \label{wsd}
\begin{tabular}{|c|c|c|c|c|c|c |c|c|}\hline\hline
&$D^0K^+$ & $D^+K^0 $ & $D^{*+}K^0$ &$D^{*0}K^+$&$D_s \eta$ & $D^*_s \eta$ & total & $\Gamma(D^*K)/\Gamma(DK)$   \\
\hline
$1^3D_1$&75 & 73.6 & 17.8&18.5 & 14 & 0.9 & 200  & 0.24\\
\hline
$2^3S_1$&5.4   & 5.6 & 9.0&9.1  & 1.7 & 0.7 & 31 &  1.65\\
\hline
\end{tabular}
\end{table}

We first assign the $D_{sJ}(2710)$ as the $2^3S_1$ and $1^3D_1$
states and calculate its decay widths. The results are listed in
Tab.~\ref{wsd}, respectively. For the assignment of the $2^3S_1$
state, the total decay width and the decay branching ratio fraction
between $D^*K$ and $DK$ channels are
\begin{eqnarray}
\Gamma\simeq 31 ~\mathrm{MeV},\ \
\frac{\Gamma(D^*K)}{\Gamma(DK)}\simeq 1.65.
\end{eqnarray}
It shows that the predicted width $\Gamma\simeq 31$ MeV is too
narrow to compare with the data, and the predicted decay ratio
$D^*K/DK\simeq 1.67$ is much larger than the measurement
$D^*K/DK\simeq 0.91\pm 0.13\pm0.12$ \cite{Aubert:2009}. The
calculations of Ref.~\cite{Zhang:2006yj} also tend to give a small
width $\Gamma\simeq 32$ MeV for the $2^3S_1$ configuration. The
predicted branching ratio fraction is also inconsistent with the
observations \cite{Aubert:2009}. In Ref.~\cite{Close:2006gr}, it is
also found that the large branching ratio fraction $D^*K/DK\simeq
3.55$ does not support the $D_{sJ}(2710)$ as a pure $2^3S_1$ state.

On the other hand, if the $D_{sJ}(2710)$ is considered as a $1^3D_1$
state, the decay width and branching ratio fraction will be
\begin{eqnarray}
\Gamma\simeq 200 ~\mathrm{MeV},\ \
\frac{\Gamma(D^*K)}{\Gamma(DK)}\simeq 0.24.
\end{eqnarray}
In this case, the branching ratio fraction
$\Gamma(D^*K)/\Gamma(DK)\simeq 0.24$ is too small though the decay
width $\Gamma\simeq 200$ MeV is roughly consistent with the upper
limit of the data \cite{jb:2007aa,Aubert:2009}. These results
suggest that either $1^3D_1$ or $2^3S_1$ is not a good assignment
for the $D_{sJ}(2710)$.

\begin{center}
\begin{figure}[ht]
\centering \epsfxsize=10 cm \epsfbox{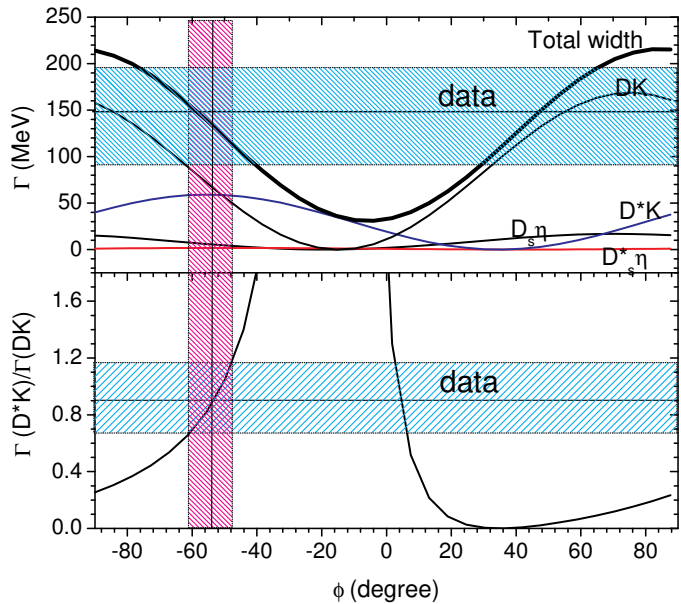} \caption{ (Color
online) The partial decay widths, total width,  and the decay
branching ratio fraction $\Gamma(D^*K)/\Gamma(D^*K)$ of
$|(SD)_1\rangle_L$ as functions of mass, respectively. The data are
from BABAR \cite{Aubert:2009}.}\label{fig-8}
\end{figure}
\end{center}

Thus, we consider the possibilities of the $D_{sJ}(2710)$ as a mixed
state of $2^3S_1$-$1^3D_1$, for which the physical states can be
expressed as~\cite{Close:2006gr}
% The $2^3S_1$-$1^3D_1$ mixing may occur by the
%spin-orbit interaction or other reasons, thus, the physical low mass
%state and its heavy partner are given by
\begin{eqnarray}
|(SD)_1\rangle_L=+\cos (\phi) |2^3S_1\rangle+\sin(\phi)|1^3D_1\rangle, \\
|(SD)'_1\rangle_H=-\sin (\phi)
|2^3S_1\rangle+\cos(\phi)|1^3D_1\rangle \ ,
\end{eqnarray}
where the physical partner in the mixing is included. Assuming that
the low mass state $|(SD)_1\rangle_L$ corresponds to the
$D_{sJ}(2710)$~\cite{Close:2006gr}, we plot the decay properties of
$|(SD)_1\rangle_L$ as functions of the mixing angle $\phi$ in
Fig.~\ref{fig-8}. It shows that with the mixing angle
$\phi\simeq(-54\pm 7)^\circ$, the decay width and branching ratio
fraction are
\begin{eqnarray}
\Gamma\simeq (133\pm 22) ~\mathrm{MeV},\ \
\frac{\Gamma(D^*K)}{\Gamma(DK)}\simeq 0.91\mp 0.25 \ ,
\end{eqnarray}
which are in a good agreement with the data
\cite{jb:2007aa,Aubert:2009}.

\begin{center}
\begin{figure}[ht]
\centering \epsfxsize=10 cm \epsfbox{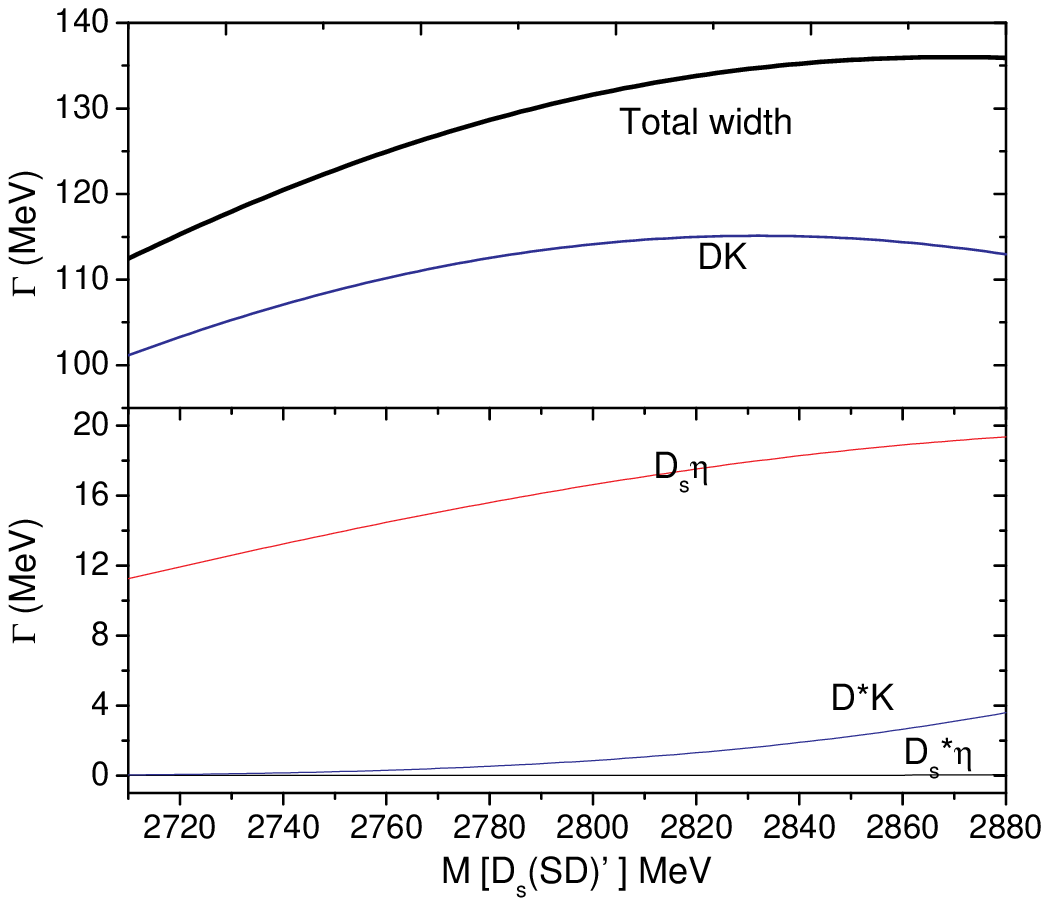} \caption{ (Color
online) The partial decay widths and total width of
$|(SD)'_1\rangle_H$ as functions of mass.}\label{fig-9}
\end{figure}
\end{center}

Following this scheme, one can examine the high-mass partner
$|(SD)'_1\rangle_H$, of which the expected mass is $\sim 2.81$
GeV~\cite{Close:2006gr}.  Taking into account the mass uncertainties
of a region $M\simeq (2.71\sim 2.88)$ GeV, we plot the
mass-dependence of the partial and total widths in Fig.~\ref{fig-9}.
It shows that the $|(SD)'_1\rangle_H$ also has a broad width $\sim
(120\pm 10)$ MeV, and the $DK$ channel is dominant over others. In
contrast, the partial width of $D_s \eta$ is also sizable, while the
$D^*_s\eta$ width is negligible. Around $M=2.81$ GeV, the predicted
branching ratio fractions are
\begin{eqnarray}
\frac{\Gamma(D_s\eta)}{\Gamma(DK)}\simeq
0.15,~\frac{\Gamma(D^*K)}{\Gamma(D_s\eta)}\simeq 0.06.
\end{eqnarray}

The above mixing scheme is consistent with Ref.~\cite{Close:2006gr}
for the low-mass state while the predicted suppression of the $D^*K$
decay mode is different from that of Ref.~\cite{Close:2006gr}. In
Ref.~\cite{Close:2006gr} a very broad high-mass state is predicted
and would dominantly decay into both $DK$ and $D^*K$. In our scheme,
the predicted decay width for $|(SD)'_1\rangle_H$ is $\sim (120\pm
10)$ MeV. As a consequence, one would expect that it should appear
in the $DK$ spectrum similar to the $D_s(2710)$ signal. Taking into
account the still undetermined mass for $|(SD)'_1\rangle_H$, one
possible explanation would be that the $|(SD)'_1\rangle_H$ mass may
be larger than $M\simeq 2.88$ GeV. If so, its total width would be
larger than we estimated above and become much broader, thus, cannot
be easily identified in the present $DK$ spectrum. Interestingly, a
recent study of the $D_s$ spectrum suggests a larger mass for the
$1^3D_1$ state~\cite{Ebert:2009ua}.

It should be noted that different methods seem to lead to different
conclusions on the $D_{sJ}(2710)$ state. In
Refs.~\cite{Colangelo:2007ds,Fazio}, both the decay width and
branching ratio fraction of the $D_{sJ}(2710)$ as the $2^3S_1$
assignment can be well explained. In Ref.~\cite{Ebert:2009ua}, the
mass calculation also suggests that the $D_{sJ}(2710)$ is $2^3S_1$.
However, the recent study of a $^3P_0$ model tends to conclude that
the $D_{sJ}(2710)$ is a mixture of $2^3S_1$ and
$1^3D_1$~\cite{Li:2009qu}. Therefore, additional information for
$D_{sJ}(2710)\to D_{s}\eta$ and $D_s^{*}\eta$, as well as a search
for the $|(SD)'_1\rangle_H$ partner in experiment would be useful
for understanding the property of the $D_{sJ}(2710)$.

\subsection{$D_{sJ}(2860)$}

The situation about the $D_{sJ}(2860)$ is  still controversial and
different solutions have been proposed in the literature. In
Ref.~\cite{vanBeveren:2006st}, the $D_{sJ}(2860)$ is assigned as a
$J^P=0^+$ state. However, the recent observation of
$D_{sJ}(2860)\rightarrow D^*K$ does not support this picture. It is
also proposed to be a $J^P=3^-$ state
\cite{Colangelo:2006rq,Zhang:2006yj,Zhong:2008kd}. However, although
the decay width and decay mode are consistent with the observation,
the predicted ratio $D^*K/DK\simeq 0.4$ is too small to compare with
the data $D^*K/DK\simeq 1.1$ \cite{Aubert:2009}.

\begin{table}[ht]
\caption{The decay widths (MeV) for the $D_{sJ}(2860)$ as $1^3D_3$,
$2^3P_2$, and $1^3F_2$ candidates.} \label{w28}
\begin{tabular}{|c|c|c|c|c|c|c |c|c|c|c}\hline\hline
       &$D^0K^+$ & $D^+K^0 $ & $D^{*+}K^0$ &$D^{*0}K^+$&$D_s \eta$ & $D^*_s \eta$
       & $DK^*$& total & $\Gamma(D^*K)/\Gamma(DK)$\\
\hline
$1^3D_3$&12.3 & 11.8 & 5 &4.7 & 1.7 & 0.3 & 0.2& 36 & 0.40 \\
\hline
$2^3P_2$&1.3 & 1.3 & 2.1 & 1.9  & 0.01 & 1.7 & 0.02& 8 & 1.53 \\
\hline
$1^3F_2$&21.9   & 21.4 & 0.1&  0.1 & 5.5 & 0.02 & 0.005 & 49 &0.005 \\
\hline
\end{tabular}
\end{table}

Since the $D_{sJ}(2860)$ is observed in both $D^*K$ and $DK$
channels, the allowed quantum numbers would be $1^3D_3$, $2^3P_2$
and $1^3F_2$. We calculate the total and partial widths for these
configurations and list the results in Tab.~\ref{w28}.

More specifically, as the $1^3D_3$ state, the predicted width and
branching ratio fraction between the $D^*K$ and $DK$ channel are
\begin{eqnarray} \Gamma\simeq 36 ~\mathrm{MeV},~
\frac{\Gamma(D^*K)} {\Gamma(DK)}\simeq 0.4.
\end{eqnarray}
The predicted ratio $\Gamma(D^*K)/\Gamma(DK)$ differs from the
measurement $D^*K/DK\simeq 1.1$ \cite{Aubert:2009} at the level of
three standard deviations, although the decay width is in agreement
with the data. Our predictions are consistent with those of Refs.
\cite{Colangelo:2006rq,Fazio}. It should be mentioned that the
QCD-motivated relativistic quark model can not well explain the mass
of $D_{sJ}(2860)$ if it is considered as the $1^3D_3$
state~\cite{Ebert:2009ua}. This could be a signal indicating the
chiral symmetry in association with the heavy quark symmetry in the
heavy-light meson transitions.

As a candidate of the $2^3P_2$ state, the decay width and branching
ratio fraction of $D_{sJ}(2860)$ are
\begin{eqnarray}
\Gamma\simeq 8 ~\mathrm{MeV},~ \frac{\Gamma(D^*K)}
{\Gamma(DK)}\simeq 1.53,
\end{eqnarray}
where both the predicted width and ratio are inconsistent with the
data. It is interesting to mention that our predicted ratio agrees
with the estimation of Ref.~\cite{eef:2009}.

If the $D_{sJ}(2860)$ is a $1^3F_2$ state, the predicted width and
branching ratio fraction are
\begin{eqnarray}
\Gamma\simeq 49 ~\mathrm{MeV},~ \frac{\Gamma(D^*K)}
{\Gamma(DK)}\simeq 0.005 \ ,
\end{eqnarray}
where the decay mode of $D^*K$ turns out to be negligible in
comparison with the $DK$ mode, and disagrees with the experimental
observation.

\begin{center}
\begin{figure}[ht]
\centering \epsfxsize=10 cm \epsfbox{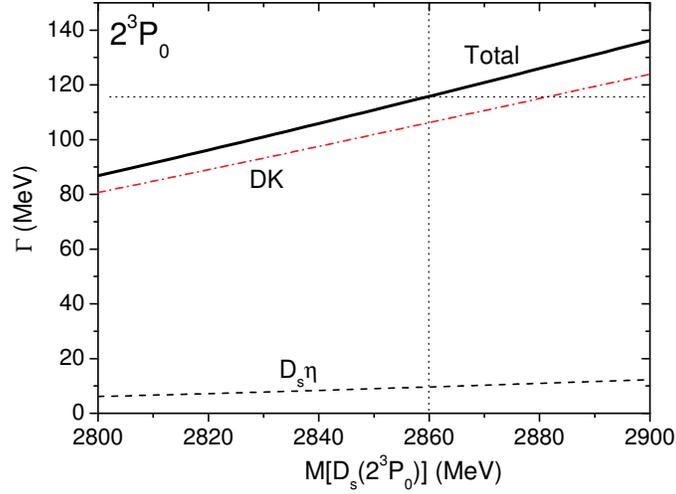} \caption{ (Color
online) The partial decay widths and total width of $2^3P_0$  as
functions of mass.}\label{fig-6}
\end{figure}
\end{center}

\begin{center}
\begin{figure}[ht]
\centering \epsfxsize=10 cm \epsfbox{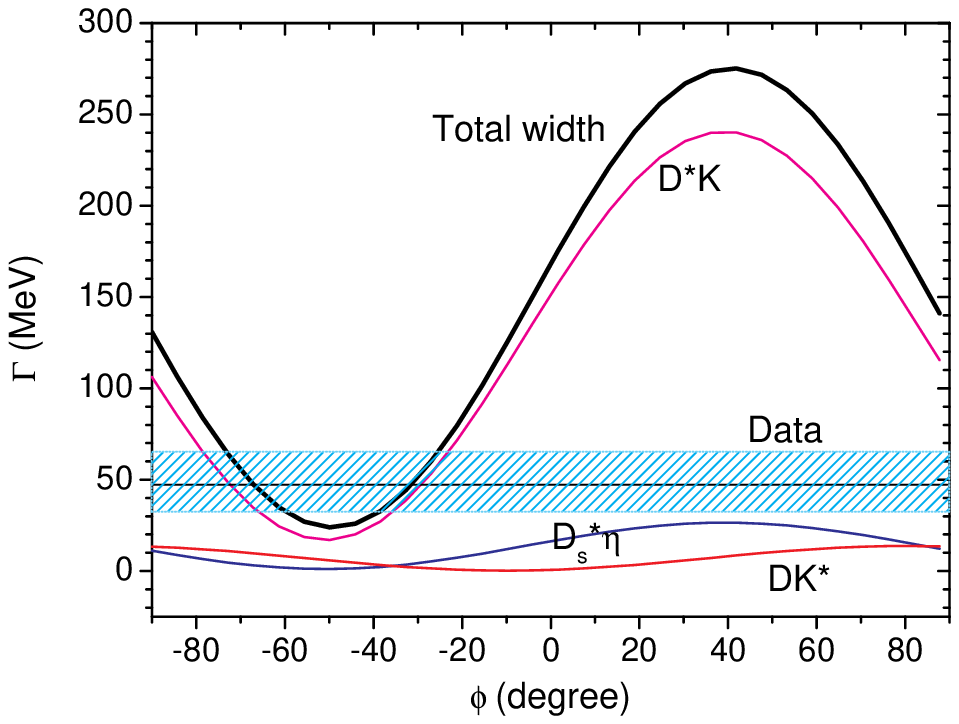} \caption{ (Color
online) The partial decay widths and total width of $|1D_2\rangle_H$
with a mass of 2860 MeV as functions of mixing angle $\phi$. The
data are from BABAR \cite{Aubert:2009,Aubert:2006mh}.}\label{fig-5}
\end{figure}
\end{center}

It can be seen from the above analysis that a simple assignment of
the $D_{sJ}(2860)$ to be a pure $2^3P_0$, $1^3D_3$, $2^3P_2$ or
$1^3F_2$ cannot well explain the data. We also point out that the
$2^3P_2$ and $1^3F_2$ mixing is unable to overcome the problem
either because of the narrow width of the $2^3P_2$ state or small
branching ratio fraction $\Gamma(D^*K)/\Gamma(DK)\simeq 0.005$ of
$1^3F_2$.

In Ref.~\cite{eef:2009}, van Beveren and Rupp recently proposed an
alternative solution that there might exist two largely overlapping
resonances at about 2.86 GeV, i.e. a radially excited tensor ($2^+$)
and a scalar ($0^+$) $c\bar{s}$ state. Following this two-state
assumption, one would expect that one state $D_{sJ_1}(2860)$
dominantly decays into $DK$, while the other one $D_{sJ_2}(2860)$
dominantly decays into $D^*K$. Both states have a mass around 2.86
GeV, and comparable width $\Gamma\sim 50$ MeV. This idea may shed
some light on the controversial issues. As follows, we shall
investigate such a possibility in our approach.

It shows that the decays of $2^3P_0$, $1^3D_3$ and $1^3F_2$ is
dominated by the $DK$ channel, while the decay of $1^3D_2$,
$1^1D_2$, $2^3P_2$ is dominated by the $D^*K$ channel. We shall
identify which states are more appropriate candidates in the
two-state scenario.

First, we analyze the states dominated by $DK$ decays, i.e.
$2^3P_0$, $1^3D_3$ and $1^3F_2$. In Fig.~\ref{fig-6} the total and
partial decay widths for the $2^3P_0$ state are revealed. It shows
that the $2^3P_0$ possesses a broad decay width $\Gamma\simeq 115$
MeV at about 2.86 GeV, which is inconsistent with the data. The
$1^3F_2$ is not considered as a good candidate of $D_{sJ_1}(2860)$
as well since its mass is expected to be larger than 3.1
GeV~\cite{Pierro,Ebert:2009ua}. Furthermore, our earlier analysis
suggests that the $D_{sJ}(3040)$ may favor a configuration of
$|2P_1\rangle_L$ such that the mass of the $1^3F_2$ should be larger
than the $P$ wave state $|2P_1\rangle_L$ as a consequence. In
contrast, we find that the $1^3D_3$ could be a good candidate for
$D_{sJ_1}(2860)$ since it is dominated by the $DK$ decay mode and
has a narrow width $\Gamma\simeq 36$ MeV. The calculation results
for the total and partial decay widths have been listed in
Tab.~\ref{w28}.

Candidates for the $D_{sJ_2}(2860)$ could be  $1^3D_2$, $1^1D_2$, or
$2^3P_2$ which dominantly decay into $D^*K$. As discussed earlier in
this section and shown in Tab.~\ref{w28}, the $2^3P_2$ is not a good
candidate since its total width is too small to compare with the
data. Nevertheless, its expected mass should be larger than 2.86 GeV
~\cite{Pierro,Ebert:2009ua}.

If the $D_{sJ_2}(2860)$ is considered as pure $1^3D_2$ or $1^1D_2$
state, their decay widths would be $\Gamma\simeq 170$ MeV and
$\Gamma\simeq 130$ MeV, respectively, which are inconsistent with
the data as well. In fact, the physical states should be the
admixtures between $1^3D_2$ and $1^1D_2$ due to the presence of the
spin-orbit interactions \cite{Godfrey:1986wj,Close:2005se,Swanson}.
Thus, the mixed states can be expressed as
\begin{eqnarray}
|1D_2\rangle_L=+\cos (\phi) |1^1D_2\rangle+\sin(\phi)|1^3D_2\rangle, \\
|1{D_2}'\rangle_H=-\sin
(\phi)|1^1D_2\rangle+\cos(\phi)|1^3D_2\rangle \ ,
\end{eqnarray}
where the subscripts $L$ and $H$ denote the low-mass and high-mass
state due to the mixing. Usually, the $|1{D_2}'\rangle_H$ have a
narrow width~\cite{Godfrey:1986wj,Close:2005se,Swanson}. We thus
consider the $|1{D_2}'\rangle_H$ as the $D_{sJ_2}(2860)$ in the
calculation. In Fig.~\ref{fig-5} the decay properties as a function
of the mixing angle are plotted. We see that around $\phi=-65^\circ$
or $\phi=-35^\circ$ the decay width is $\Gamma\simeq 40$ MeV, which
is compatible with the observation, and the decay mode is dominated
by the $D^*K$. With $\phi=-35^\circ$, the corresponding decay
branching ratio fractions are
\begin{eqnarray}
\frac{\Gamma(D^*K)} {\Gamma(D^*_s\eta)}\simeq 1.2,
~\frac{\Gamma(D^*K)} {\Gamma(DK^*)}\simeq 13 \ ,
\end{eqnarray}
which fit in the experimental data quite well. This result turns out
to support the $|1{D_2}'\rangle_H$ to be a candidates of
$D_{sJ_2}(2860)$ in the two-state scenario. In the range of
$\phi=-65^\circ \sim -35^\circ$ the partial widths do not change
drastically with the mixing angle. In contrast, the suggested value
is consistent with that ($\phi=-50.7^\circ$) obtained in the heavy
quark effective theory
~\cite{Godfrey:1986wj,Close:2005se,Swanson,Ebert:2009ua}.

In brief, it seems likely that the abnormal property with the
$D_{sJ}(2860)$ arises from two overlapping resonances with the same
mass but different decay modes. One is $1^3D_3$ and the other is
$|1{D_2}'\rangle_H$ from the $1^3D_2$ and $1^1D_2$ mixing. The
$1^3D_3$ state mainly decays into $DK$ and the $|1{D_2}'\rangle_H$
into $D^*K$. With these two largely overlapping resonances at about
2.86 GeV, we can understand both the observed decay widths and
branching ratio fractions of the $D_{sJ}(2860)$. It shows that the
$1^3D_3$ has a sizable partial width in the $D_s\eta$ channel, while
the $|1{D_2}'\rangle_H\to D^*_s\eta$ also turns out to be
measurable. Further measurements of $\Gamma(D^*_s\eta)/\Gamma(D^*K)$
and $\Gamma(DK)/\Gamma(D_s\eta)$ may be able to distinguish the
$1^3D_3$ and $|1{D_2}'\rangle_H$ and test the two-state scenario in
experiment.

\begin{center}
\begin{figure}[ht]
\centering \epsfxsize=10 cm \epsfbox{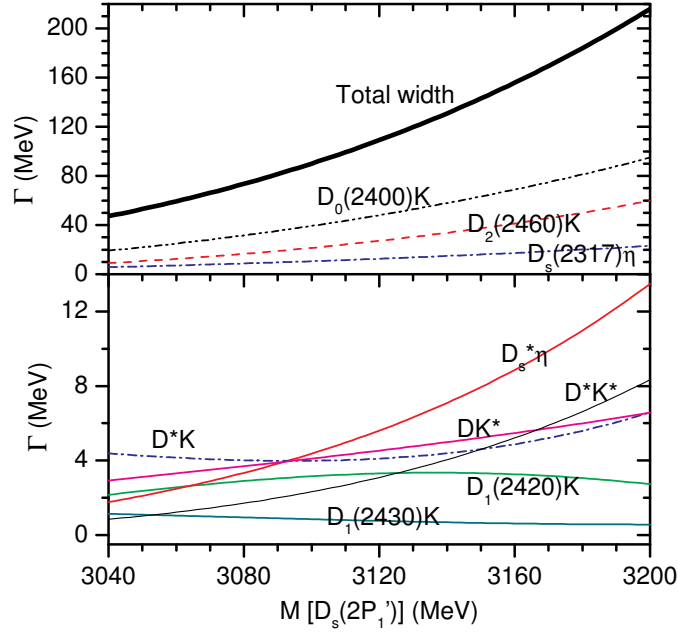} \caption{ (Color
online) The partial decay widths and total width of the
$|2P'_1\rangle_H$ state as functions of mass. }\label{fig-3}
\end{figure}
\end{center}

\subsection{$D_{sJ}(2|P'_1\rangle_H )$, $D_{sJ}(2^3P_0)$ and $D_{sJ}(2^3P_2)$}

In this subsection we discuss the implications of other states
following the consequence of the assignments for the $D_{sJ}(3040)$,
$D_{sJ}(2710)$ and $D_{sJ}(2860)$.

Since the $D_{sJ}(3040)$ seems to favor a $P$ wave with $J^P=1^+$
($|2P_1\rangle_L$), experimental evidences for the other $P$ waves,
$D_{sJ}(2|P'_1\rangle_H )$, $D_{sJ}(2^3P_0)$ and $D_{sJ}(2^3P_2)$,
would be important to establish the spectrum. In particular, its
high-mass partner $|2P'_1\rangle_H$ should be searched in
experiments. Supposing that the $|2P'_1\rangle_H$ has a mass in the
range of $(3.04\sim 3.2)$ GeV, we plot  in Fig.~\ref{fig-3} the
decay widths as functions of the mass with the mixing angle
$\phi=-50^\circ$ fixed by $D_{sJ}(3040)$. It shows that the
$|2P'_1\rangle_H$ width is indeed relative narrower around  $M=
3.04$ GeV, although we should note that the decay width increases
fast with the increasing mass. The decay channels, $D_0(2400)K$,
$D_2(2460)K$ and $D_s(2317)\eta$, are predicted to be the dominant
ones, which can be investigated in experiments. In contrast, the
$D^*K$ channel plays a less important role in the decays.

We further study the $D_{sJ}(2^3P_0)$ in detail here. The decay
widths as a function of the possible mass range $M=(2.8\sim 2.9)$
GeV are plotted in Fig.~\ref{fig-6}. In this range the total decay
width is $\Gamma\simeq (90\sim 140)$ MeV, and increases with the
increasing mass. It shows that the $DK$ channel dominates its
decays. Taking the mass of the $D_{sJ}(2^3P_0)$ as $(2.82\sim 2.84)$
GeV~\cite{Close:2006gr,Matsuki}, the total width and branching ratio
fractions between $D_s\eta$ and $DK$ are
\begin{eqnarray}
\Gamma\simeq (101\pm 5) ~\mathrm{MeV},~ \frac{\Gamma(D_s\eta)}
{\Gamma(DK)}\simeq 0.08.
\end{eqnarray}

It should be pointed out that the decay properties of $2^3P_0$ are
similar to those of $|(SD)'_1\rangle_H$ in the mass range $M< 2.9$
GeV (see Fig.~\ref{fig-9} and Fig.~\ref{fig-6}). Both of them have
comparable decay widths $\Gamma\sim 100$ MeV, and mainly decay into
$DK$. To distinguish them from each other, the measurements of their
decay ratio $\Gamma(D_s\eta)/\Gamma(DK)$ are important. We also note
that a recent calculation suggests a larger mass of $M\simeq 3.054$
GeV for $2^3P_0$~\cite{Ebert:2009ua}. As a consequence of this
scenario, its total decay width would become much broader than we
estimated above. Thus, it may not be easily isolated in experiment.

\begin{center}
\begin{figure}[ht]
\centering \epsfxsize=10 cm \epsfbox{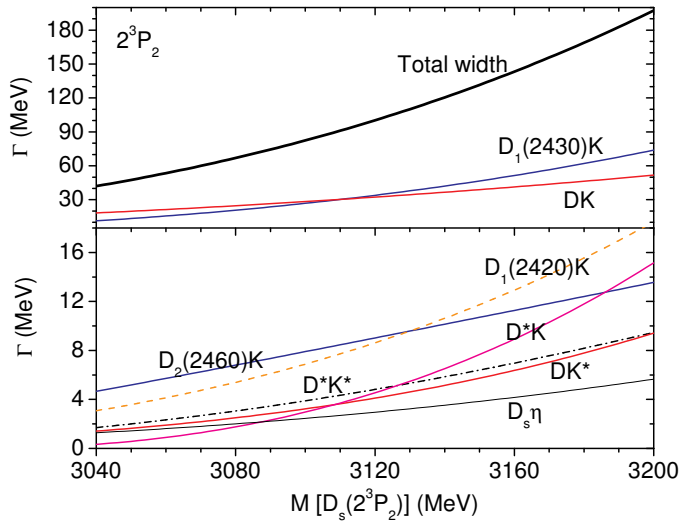} \caption{ (Color
online) The partial decay widths and total width of $2^3P_2$ as
functions of mass.}\label{fig-7}
\end{figure}
\end{center}

As discussed earlier the $D_{sJ}(2860)$ does not favor the
assignment of $2^3P_2$. Thus, we investigate its decay properties
and implications of experimental measurement. We also plot its total
and partial decay widths as functions of the mass in the possible
range $M=(3.04\sim 3.2)$ GeV in Fig.~\ref{fig-7}. If $2^3P_2$ has a
mass larger than 3.04 GeV, decay channels, $DK$, $D^*K$, $DK^*$,
$D^*K^*$, $D^*_s\eta$, $D_s\phi$,  $D_s\eta$, $D_1(2430)K$,
$D_1(2420)K$, $D_2(2460)K$, $D_s(2460)\eta$, will open in which
$D_1(2430)K$ and $DK$ channels are dominant. In Fig.~\ref{fig-7}, we
do not show the results for the $D^*_s\eta$, $D_s\phi$ and
$D_s(2460)\eta$ channels since they are negligibly small ($< 1$
MeV). If we adopt the mass $\sim 3.15$ GeV as predicted by Refs.
\cite{Matsuki,Ebert:2009ua,Pierro}, the predicted width is
$\Gamma\simeq 140$ MeV, and the relative decay strengths are
$DK:D^*K:D_1(2430)K:D_1(2420)K:D_2(2460)K:DK^*:
D^*K^*:D_s\eta\simeq41:9:50:13:11:6:7:4$. It suggests that the $DK$,
$D_1(2430)K$, $D_1(2420)K$ channels may be the optimal ones for
searching for the $D_{sJ}(2^3P_2)$ state in experiment.

\subsection{Sensitivity to the harmonic oscillator parameter}

It should be mentioned that model-dependent feature of our model
arises from the simple treatment of harmonic oscillator potential
for the heavy-light quark system. Therefore, uncertainties with the
theoretical results are present in the choice of the quark model
parameter values. The most important parameter in our model should
be the harmonic oscillator strength $\beta$, which controls the size
effect or coupling form factor from the convolution of the
heavy-light meson wavefunctions. The commonly adopted range of this
quantity is $\beta=(0.4\sim 0.5)$ GeV, and we apply $\beta=0.45$ GeV
in the above calculations.

\begin{center}
\begin{figure}[ht]
\centering \epsfxsize=9 cm \epsfbox{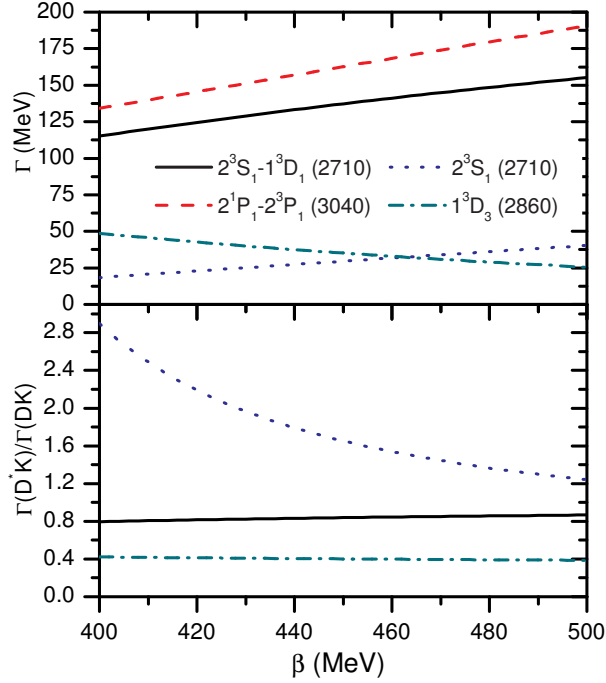} \caption{ (Color
online) The decay widths (upper panel) and ratios (lower panel) of
different configuration assignments as a function of $\beta$: the
solid lines are for the mixed state $|(SD)\rangle_L$ of
$2^3S_1$-$1^3D_1$ with a mass of 2710 MeV and mixing angle
$-55^\circ$; the dotted lines are for $2^3S_1$ with a mass of 2710
MeV; the dot-dashed lines for $1^3D_3$ with a mass of 2860 MeV; and
the dashed line in the upper panel is for the mixed state
$|2{P_1}\rangle_L$ of $2^1P_1$-$2^3P_1$ with a mass of 3040 MeV and
mixing angle $-50^\circ$. }\label{fig-11}
\end{figure}
\end{center}

In order to examine the sensitivity of the calculation results to
$\beta$, we plot the decay widths and ratios of $2^3S_1$, $1^3D_3$,
mixed state $|(SD)\rangle_L$ of $2^3S_1$-$1^3D_1$,  and mixed state
$|2{P_1}\rangle_L$ of $2^1P_1$-$2^3P_1$ as a function of $\beta$ in
Fig.~\ref{fig-11}. It shows that the decay widths of these excited
$D_s$ states exhibit some sensitivities to the parameter $\beta$.
Within the range of $\beta=(0.45\pm0.05)$ GeV, about $30\%$
uncertainties of the decay widths would be expected. This is a
typical order of accuracy for the constituent quark model, and can
be regarded as reasonable.

The ratio $\Gamma(D^*K)/\Gamma(DK)$ appears to behave differently.
For the $2^3S_1$, the sensitivity of the ratio to $\beta$ is
apparent. In contrast, the ratios of $|(SD)\rangle_L$ and $1^3D_3$
are quite insensitive to $\beta$. The ratio
$\Gamma(D^*K)/\Gamma(DK)$ is not shown for $|2{P_1}\rangle_L$ since
its decay into $DK$ is forbidden.

In brief, although the harmonic oscillator parameter $\beta$ can
bring some uncertainties to the final results, within the range of
$\beta=(0.4\sim 0.5)$ GeV, our major conclusions will still hold.

\section{summary}\label{suma}

In this work we investigate the strong decays of several newly
observed charmed mesons in a constituent quark model with effective
Lagrangians for the quark-meson interactions. The decay amplitudes
are extracted for light pseudoscalar meson or vector meson
productions via axial or vector current conservation between the
quark-level and hadronic level couplings. The quark-meson couplings
can then be determined by independent measurements such as meson
photoproduction and meson-baryon scatterings.

We find that the new state $D_{sJ}(3040)$ can be identified as the
low mass physical state $|2{P_1}\rangle_L$ from the
$D_{s}(2^1P_1)$-$D_{s}(2^3P_1)$ mixing with a mixing angle
$\phi\simeq-(40\pm 12)^\circ$. Further experimental search for decay
modes of $D_1(2430)K$, $D_2(2460)K$, $D_0(2400)K$, $DK^*$, and
$D_s^*\eta$ should be able to disentangle its property and test our
model predictions.

The $D_{sJ}(2710)$ seems to favor a low mass physical state
$|(SD)\rangle_L$ from the $2^3S_1$-$1^3D_1$ mixing with a mixing
angle $\phi\simeq (-54\pm 7)^\circ$. Both the ratio and width are in
a good agreement with the data. The decay properties of its heavy
partner $|(SD)'_1\rangle_H$ are also discussed. It has a broad width
$\Gamma\simeq (110\sim 140)$ MeV at the $2.8$ GeV mass region, and
dominated by the $DK$ mode. We also point out that the
$|(SD)'_1\rangle_H$ state may be searched in the $DK$ spectrum as
the $D_{sJ}(2710)$ if its mass is $\sim 2.8$ GeV. Whether the
present data have contained its signal could be a crucial criteria
for various model predictions.

The $D_{sJ}(2860)$ cannot be easily explained by a single
configuration of $2^3P_0$, $2^3P_2$, $1^3F_2$ or $1^3D_3$. To
overcome this problem we follow the proposal of a two-state picture
by Ref.~\cite{eef:2009} and assume that two narrow resonances may
have been observed around 2.86 GeV with a width $\Gamma\simeq
(40\sim 50)$ MeV. It shows that one resonance seems to be the
$1^3D_3$, which mainly decays into $DK$. The other resonance could
be the $|1{D_2}'\rangle_H$, which is the high-mass state from the
$1^1D_2$-$1^3D_2$ mixing, and dominantly decays into $D^*K$. Further
theoretical and experimental efforts are needed to disentangle the
mysterious properties about this state.

We also study the implications arising from the assignments for
those observed resonances, e.g. their partner states in the mixing.
In particular, if the $D_{sJ}(3040)$ is indeed a $P$-wave state
$|2{P_1}\rangle_L$, the other three $P$-wave states
$D_{sJ}(|2P'_1\rangle_H )$, $D_{sJ}(2^3P_0)$ and $D_{sJ}(2^3P_2)$
may also have measurable effects in experiment. Their strong decay
properties are predicted, which could be useful for future
experimental studies.

%%%%%%%%%%%%%%%%%%%%%%%%%%%%%%%%%%%%%%%%%%%%%%%%%%%%%%%%%%%%%%%%%%%%%

\section*{  Acknowledgements }

This work is supported, in part, by the National Natural Science
Foundation of China (Grants 10675131 and 10775145), Chinese Academy
of Sciences (KJCX3-SYW-N2), and Ministry of Science and Technology
of China (2009CB825200).
%\appendix
%%%%%%%%%%%%%%%%%%%%%%%%%%%%%%%%%%%%%%%%%%%%%%%%%%%%%%%%%%%%%%%%%%555

\end{document}